\documentclass[preprint,onecolumn]{revtex4}
\usepackage{amsmath}
\usepackage{amssymb}
\usepackage{graphicx}
\usepackage{subfigure}

\begin{document}

\title{Molecule States in a Gate Tunable Graphene Double Quantum Dot}
\author{Lin-Jun Wang}
\author{Hai-Ou Li}
\author{Zhan Su}
\author{Tao Tu}
\email{tutao@ustc.edu.cn}
\author{Gang Cao}
\author{Cheng Zhou}
\author{Xiao-Jie Hao}
\author{Guang-Can Guo}
\author{Guo-Ping Guo}
\email{gpguo@ustc.edu.cn}
\affiliation{ Key Laboratory of Quantum Information, University of Science and
Technology of China, Chinese Academy of Sciences, Hefei 230026, People's
Republic of China}

\date{\today }

\begin{abstract}
We have measured a graphene double quantum dot device with multiple
electrostatic gates that are used to enhance control to investigate it. At
low temperatures the transport measurements reveal honeycomb charge
stability diagrams which can be tuned from weak to strong interdot tunnel
coupling regimes. We precisely extract a large interdot tunnel coupling
strength for this system allowing for the observation of tunnel-coupled
molecular states extending over the whole double dot. This clean, highly
controllable system serves as an essential building block for quantum
devices in a nuclear-spin-free world.
\end{abstract}

\date{\today }
\maketitle

%%%%%%%%%%%%%%%%%%%%%%%%%%%%%%%%%%main txt%%%%%%%%%%%%%%%%%%%%%%%%%%%%%%%%%%%%

Graphene exhibits novel electrical properties and offers substantial
potential as building blocks of nanodevices owing to its unique
two-dimensional structure \cite{Geim2007,Geim2009,Ensslin2009}. Besides
being a promising candidate for high performance electronic devices,
graphene may also be used in the field of quantum computation, which
involves exploration of the extra degrees of freedom provided by electron
spin, in addition to those due to electron charge. During the past few years
significant progress has been achieved in implementation of electron spin
qubits in semiconductor quantum dots \cite{Petta2007,Hanson2008}. To realize
quantum computation, the effects of interactions between qubits and their
environment must be minimized \cite{Loss2009}. Because of the weak
spin-orbit coupling and largely eliminated hyperfine interaction in
graphene, it is highly desirable to coherently control the spin degree of
freedom in graphene nanostructures for quantum computation \cite%
{Loss2007,Guo2009}. Recently, there was a striking advance on experimental
production of graphene single or double quantum dots \cite%
{Geim2008,Stampfer2008,Molitor2010,Liu2010}, which is an important first
step towards such promise.

Here we report an experimental demonstration and electrical transport
measurement in a tunable graphene double quantum dot device. Depending on
the strength of the interdot coupling, the device can form atomiclike states
on the individual dots (weak tunnel coupling) or molecularlike states of the
two dots (strong tunnel coupling). We also extract the interdot tunnel
coupling by identifying and characterizing the molecule states with wave
functions extending over the whole graphene double dot. The result implies
that this artificial graphene device may be useful for implementing
two-electron spin manipulation.

\begin{figure}[tbp]
\scalebox{0.9}{\includegraphics*[width=0.9\linewidth]{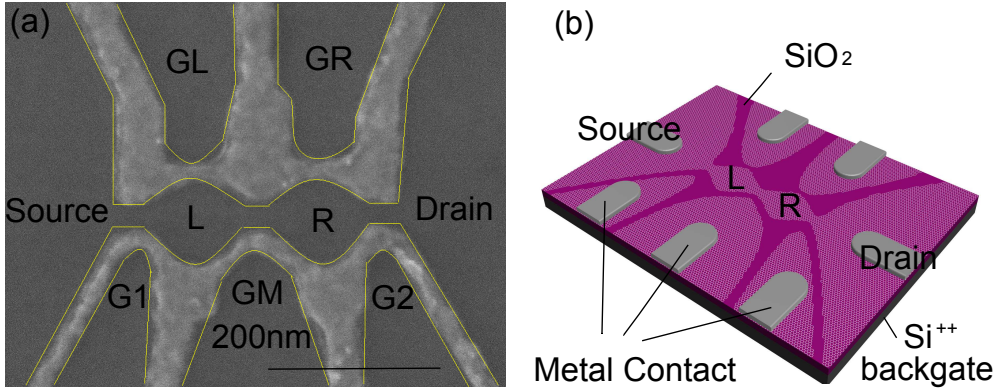}}
\caption{(a) Scanning electron microscope image of the structure of the
designed multiple gated sample studied in this work. The double quantum dot
have two isolated central islands of diameter $100$ nm in series, connected
by $20\times 20$ nm tunneling barriers to source and drain contacts (S and
D) and $30\times 20$ nm tunneling barrier with each other. These gates are
labeled by G1, GL, GM, GR, G2, in which gate GM, G1 and G2, are used to
control the coupling barriers between the dots as well as the leads. Gates
GL and GR are used to control and adjust the energy level of each dot. (b)
Schematic of a representative device.}
\end{figure}

The graphene flakes were produced by mechanical cleaving of graphite
crystallites by scotch tape and then were transferred to a highly doped Si
substrate with a $100$ nm SiO$_{2}$ top layer. Thin flakes were found by
optical microscopy and single layer graphene flakes were selected by the
Raman spectroscopy measurement. Next, a layer of poly(methyl methacrylate)
(PMMA) was exposed by standard electron beam lithography (EBL) to form a
designed pattern. The unprotected areas were carved by oxygen reactive ion
etching. We used the standard EBL and lift off technique to make the ohmic
contact (Ti/Au) on the present graphene device. A scanning electron
microscope image of our defined sample structure with double quantum dot is
shown in Figure 1. The double quantum dot has two isolated central island of
diameter $100$ nm in series, connected by $20\times 20$ nm narrow
constriction to source and drain contacts (S and D electrodes) and $30\times
20$ nm narrow constriction with each other. These constrictions are expected
to act as tunnel barriers due to the quantum size effect. In addition, the
highly P-doped Si substrate is used as a back gate and five lateral side
gates, labeled the left gate G1, right gate G2, center gate GM and GL(R),
which are expected for local control. All of side gates are effective, in
which gates GL, GR and G2 have very good effect on two dots and middle
barrier, while gates G1 and GM have weak effect on those. The device was
first immersed into a liquid helium storage dewar at $4.2$ K to test the
functionality of the gates. The experiment was carried out in a He3 cryostat
equipped with filtered wiring and low-noise electronics at the base
temperature of $300$ mK. In the measurement, we employed the standard AC
lock-in technique with an excitation voltage 20 {$\mu $V at }${11.3}${\ Hz. }

\begin{figure}[tbp]
\subfigure[] {\includegraphics[width=0.45\columnwidth]{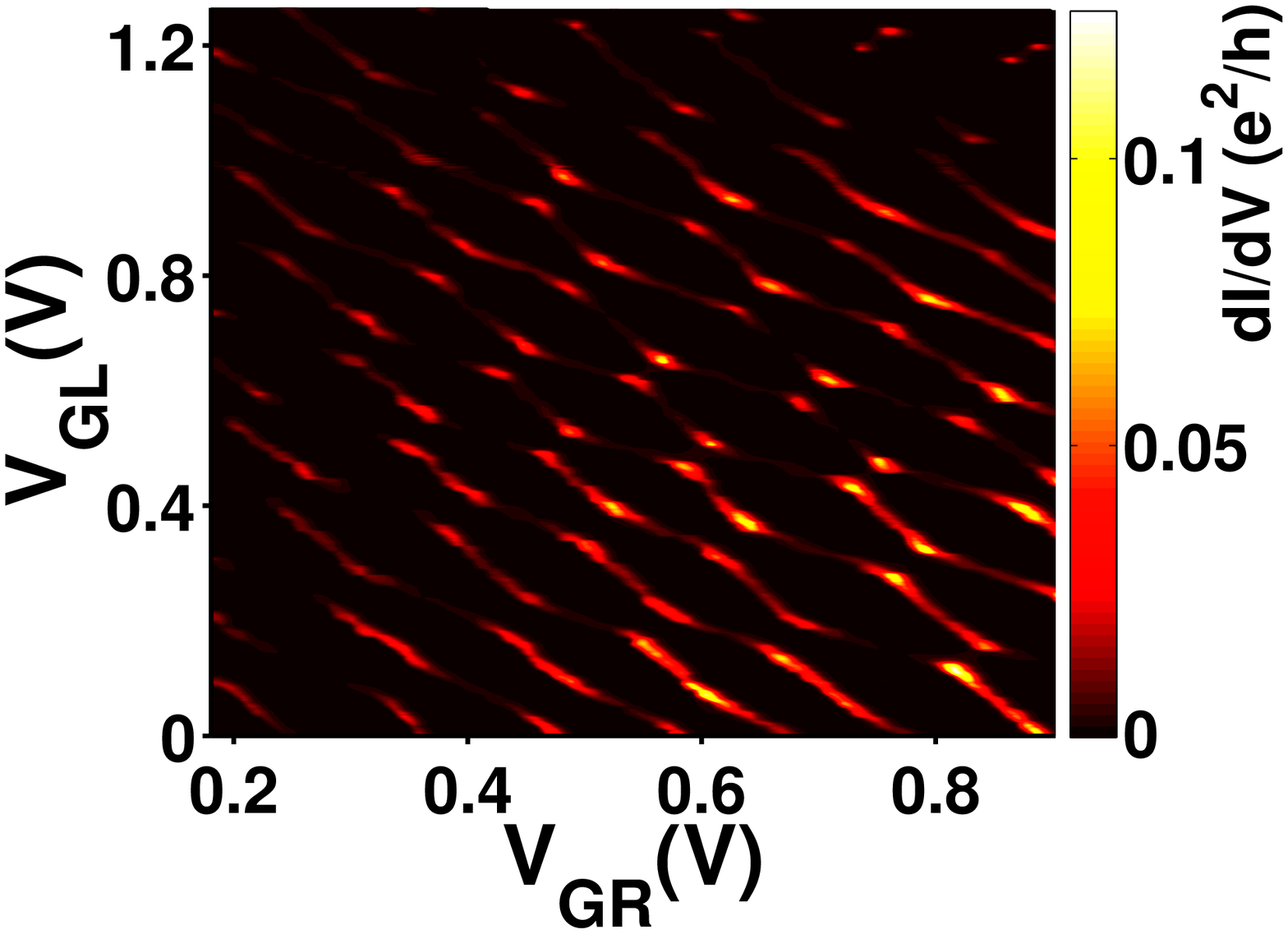}} %
\subfigure[] {\includegraphics[width=0.45\columnwidth]{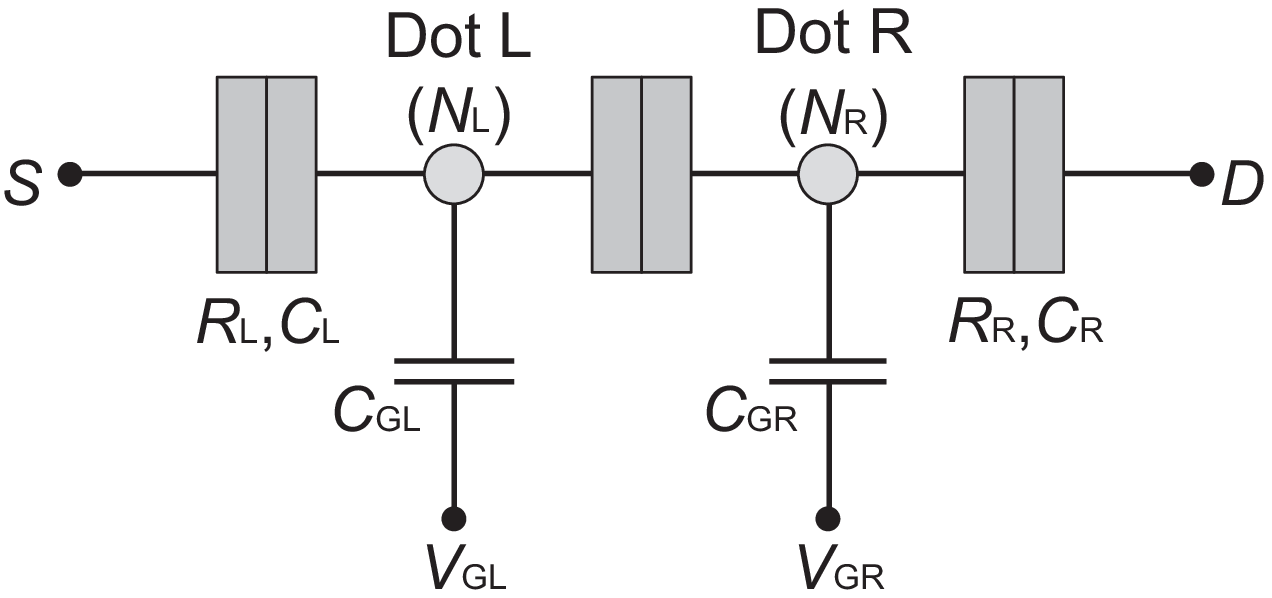}} %
\subfigure[] {\includegraphics[width=0.45\columnwidth]{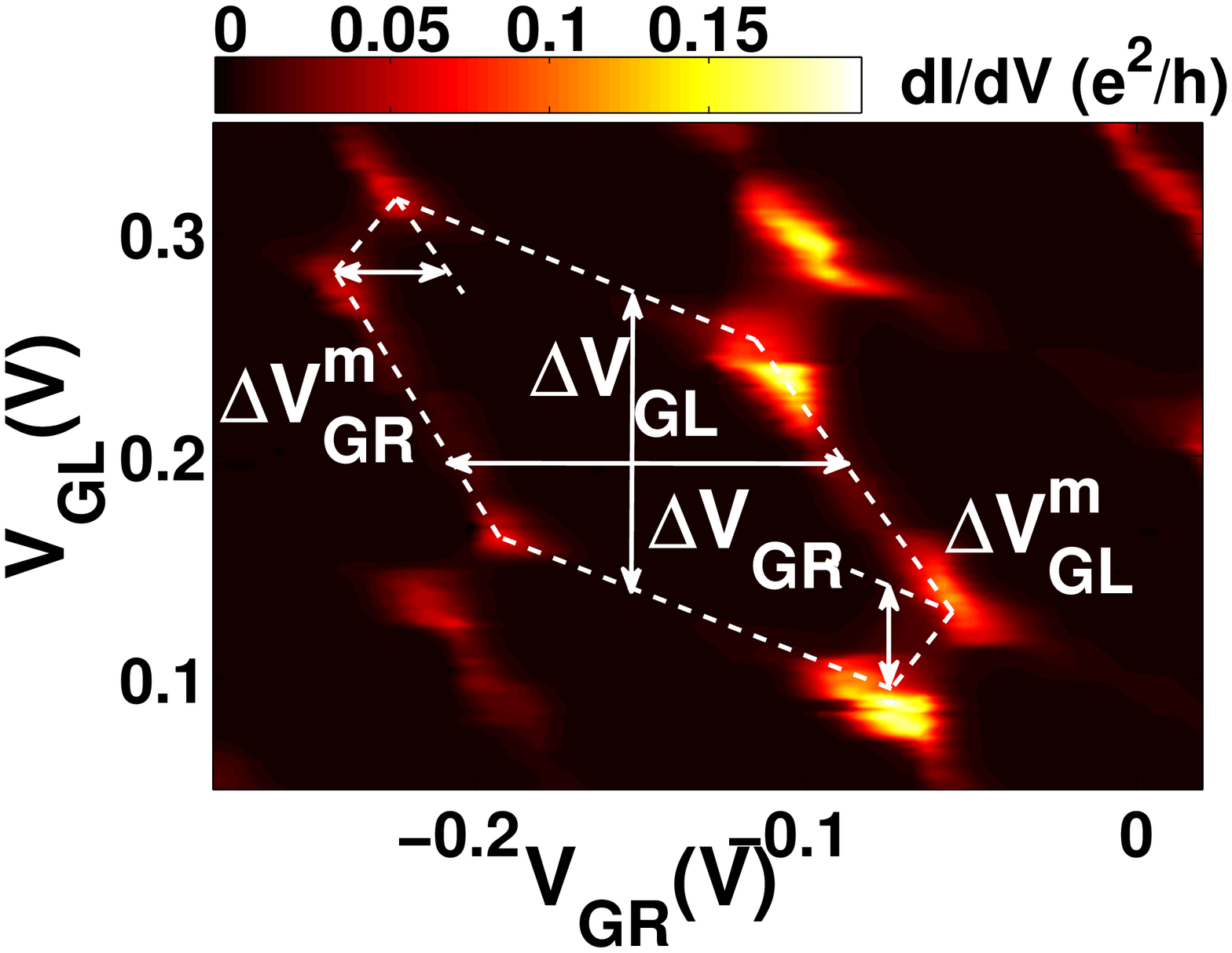}} %
\subfigure[] {\includegraphics[width=0.45\columnwidth]{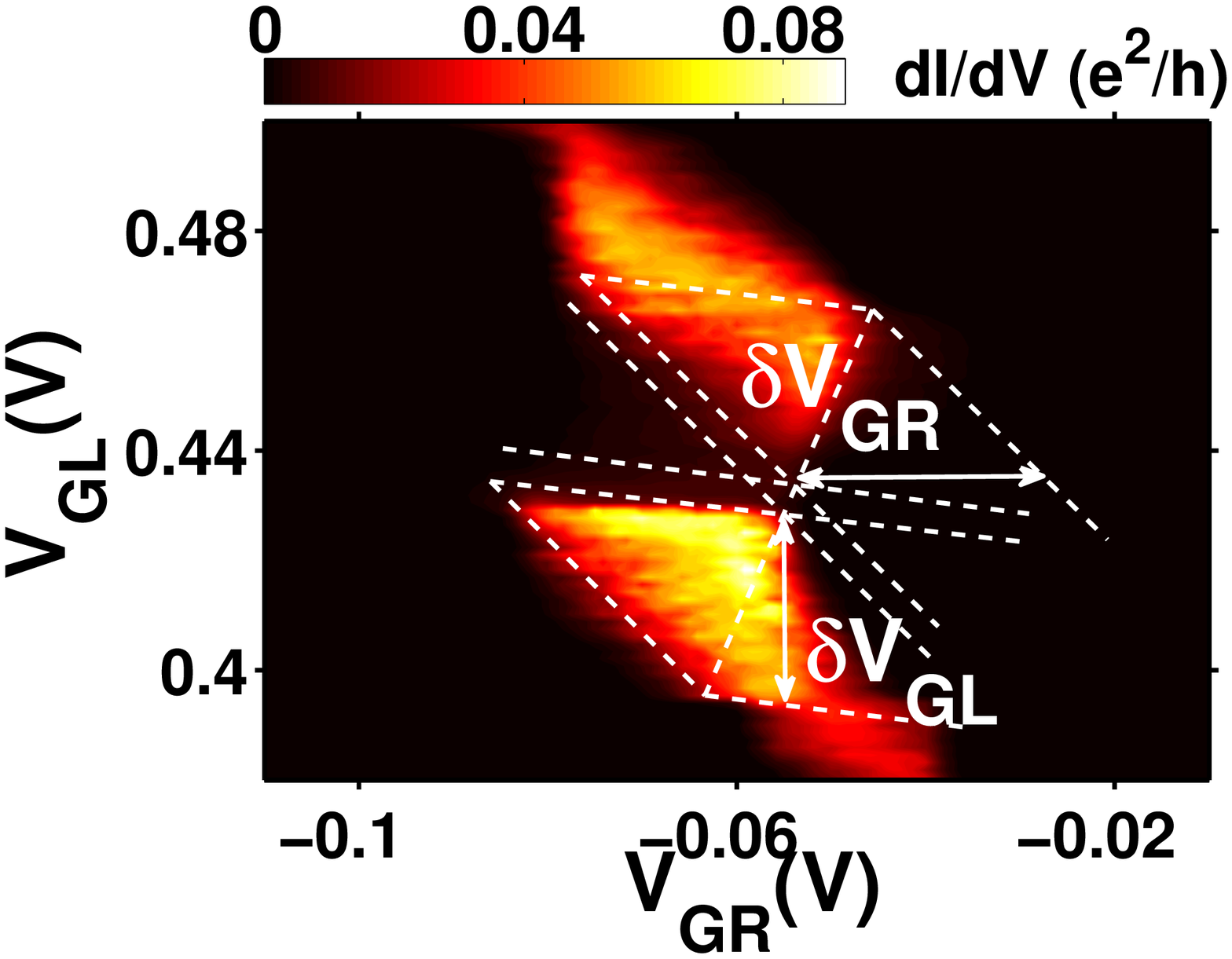}}
\caption{(a) Colorscale plot of the differential conductance versus voltage
applied on gate~GL ($V_{GL}$) and gate~GR ($V_{GR}$) at {$V_{sd}=20$%
\thinspace $\protect\mu $V}, {$V_{G1}=0$\thinspace $\protect\mu $V}, {$%
V_{GM}=0$\thinspace $\protect\mu $V} and {$V_{G2}=0$\thinspace $\protect\mu $%
V}. The honeycomb pattern we got stands for the typical charge stability
diagram of coupled double quantum dots. (b) Pure capacitance model of a
graphene double dot system. Zoom-in of a honeycomb structure (c) and a
vertex pair (d) at {$V_{sd}=900$\thinspace $\protect\mu $V}.}
\end{figure}

Fig. 2(a) displays the differential conductance through the graphene double
quantum dot circuit as a function of gate voltages $V_{GL}$ and $V_{GR}$.
Here the measurement was recorded at {$V_{sd}=20$\thinspace $\mu $V}, {$%
V_{G1}=0$\thinspace $\mu $V}, {$V_{GM}=0$\thinspace $\mu $V}, {$V_{G2}=0$%
\thinspace $\mu $V and $V_{bg}=2.5$\thinspace V}. The honeycomb pattern is
clearly visible and uniforms over many times. Each cell of the honeycomb
corresponds to a well-defined charge configuration $(N_{L},N_{R})$ in the
nearly independent dots, where $N_{L}$ and $N_{R}$ denote the number of
electrons on the left and right dot, respectively. The conductance is large
at the vertices, where the electrochemical potentials in both dots are
aligned with each other and the Fermi energy in the leads and resonant
sequential tunneling is available. These vertices are connected by faint
lines of much smaller conductance along the edges of the honeycomb cells. At
these lines, the energy level in one dot is aligned with the electrochemical
potential in the corresponding lead and inelastic cotunneling processes
occur. The observed honeycomb pattern resembles the charge stability diagram
found for weakly coupled GaAs double quantum dot \cite{VanderWiel2003}. Such
similarities indicate that graphene quantum dot devices will continue to
share features with well-studied semiconductor quantum dot systems.

\begin{figure}[tbp]
\subfigure[] {\includegraphics[width=0.40\columnwidth]{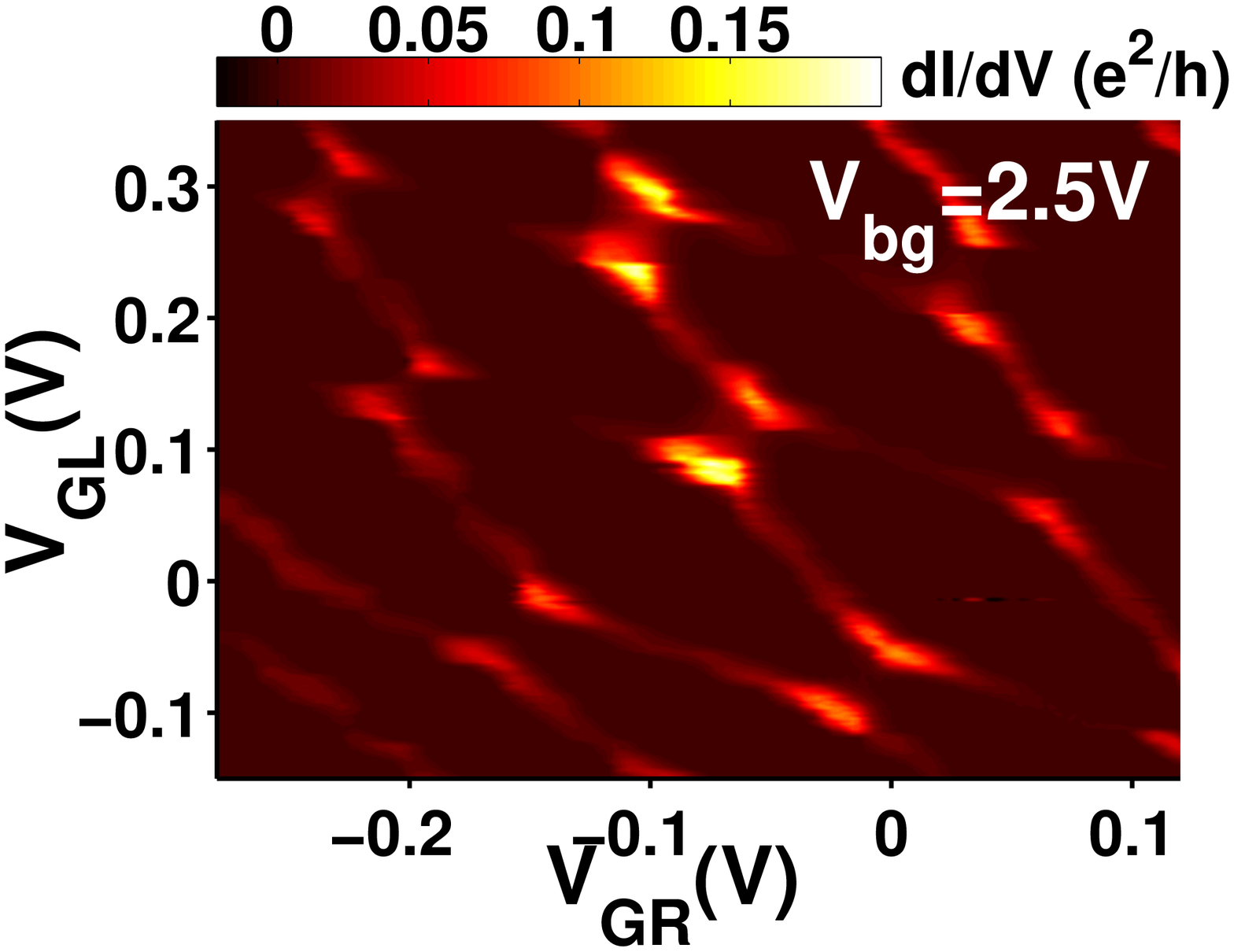}} %
\subfigure[] {\includegraphics[width=0.40\columnwidth]{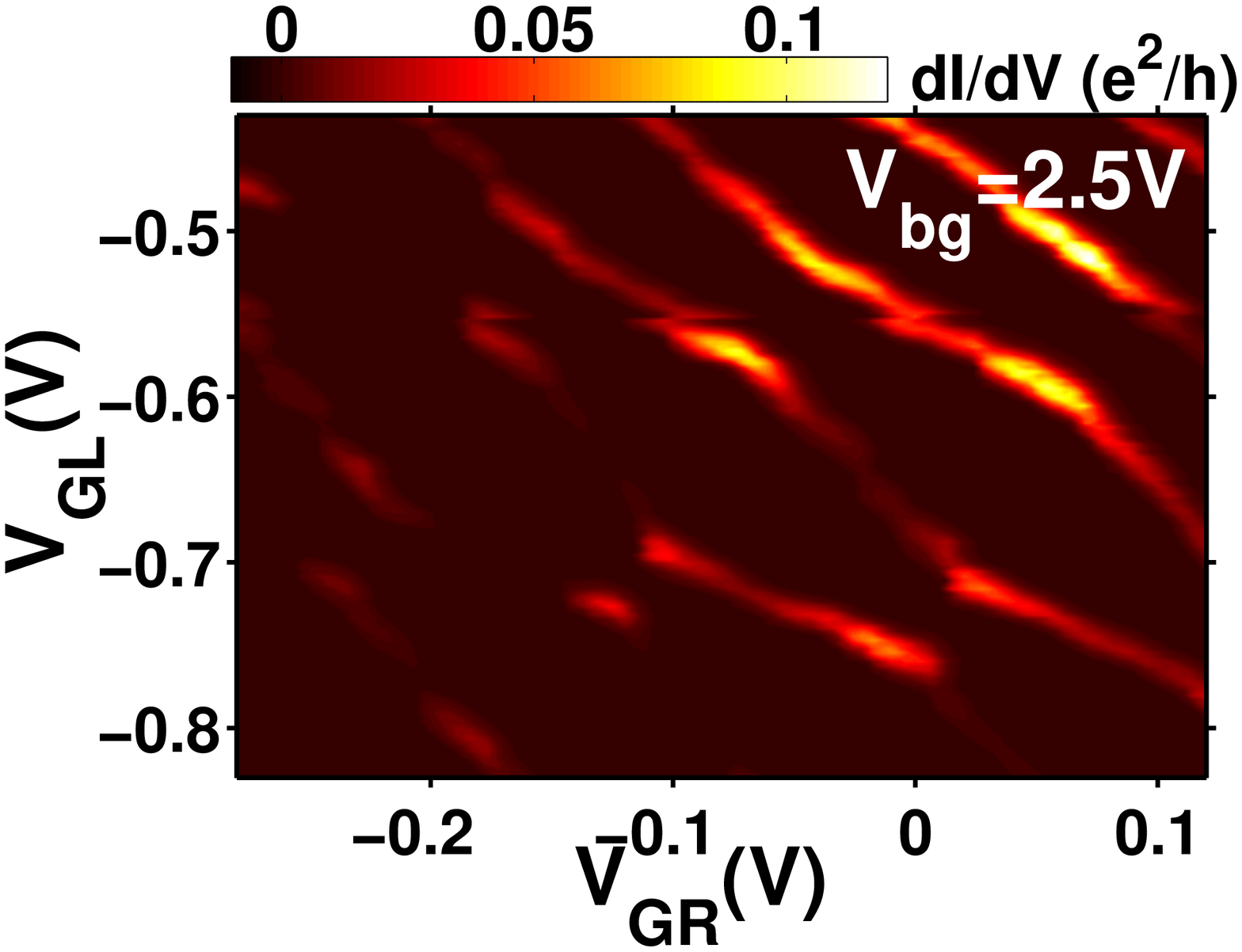}} %
\subfigure[] {\includegraphics[width=0.40\columnwidth]{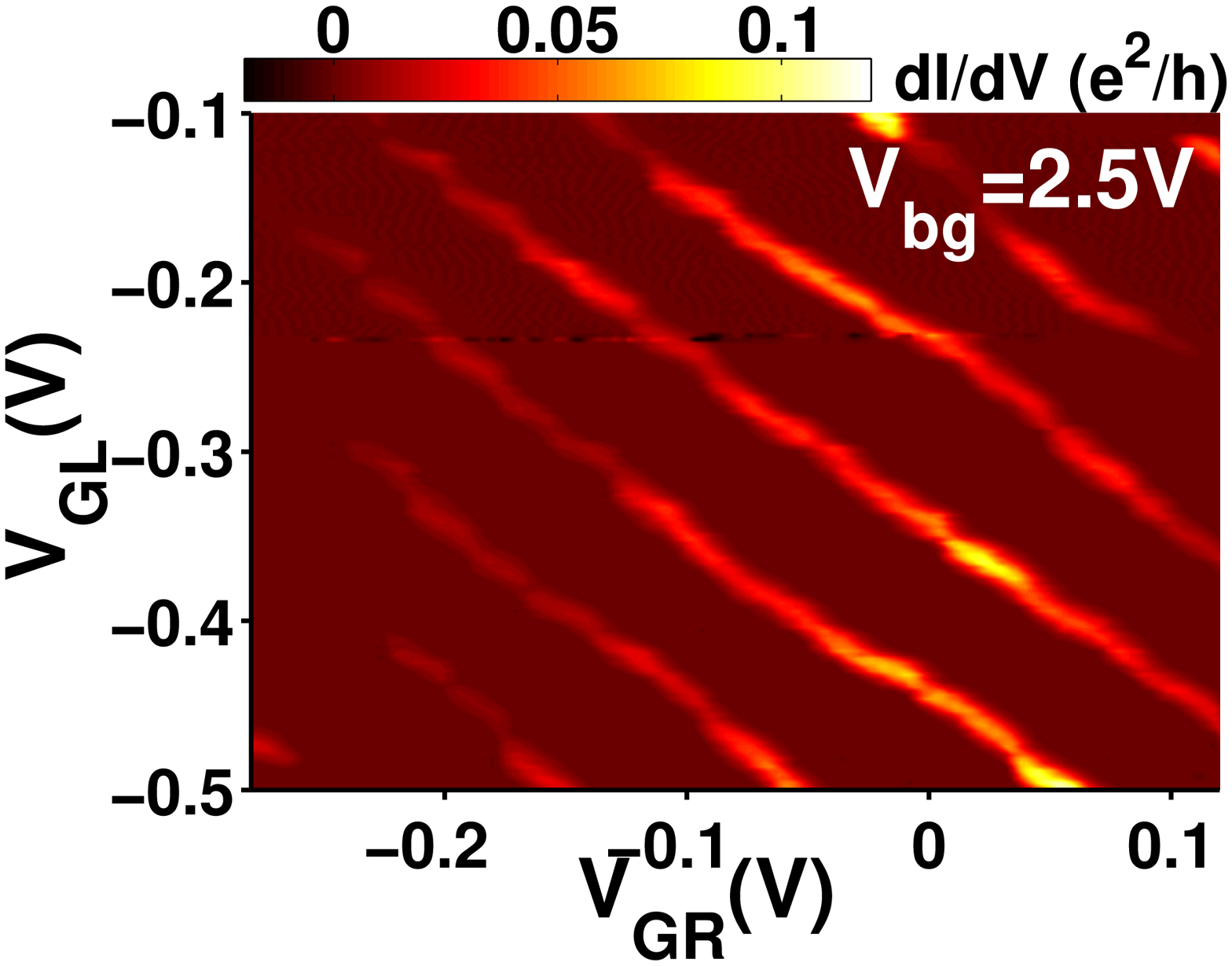}} %
\subfigure[] {\includegraphics[width=0.40\columnwidth]{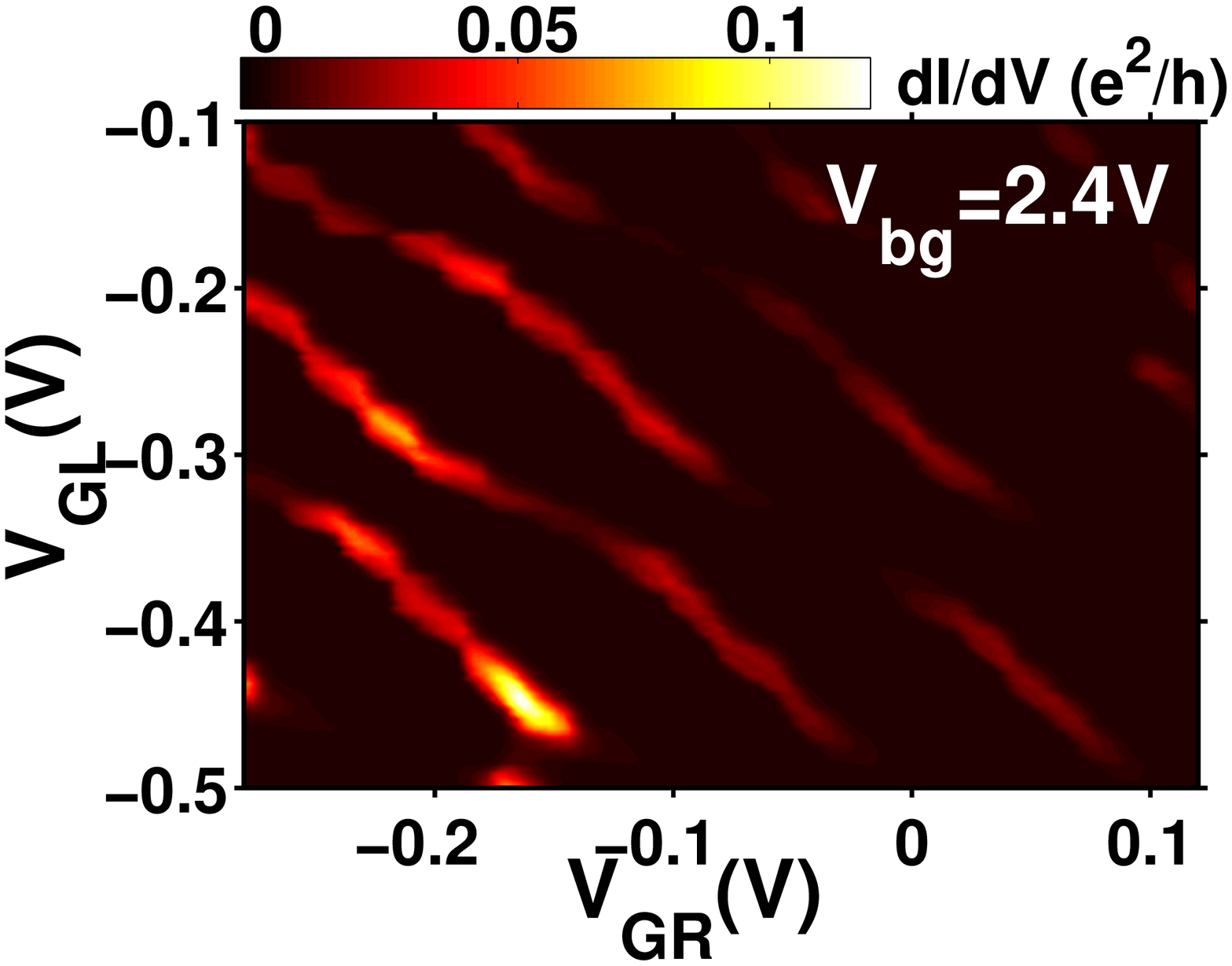}} %
\subfigure[] {\includegraphics[width=0.40\columnwidth]{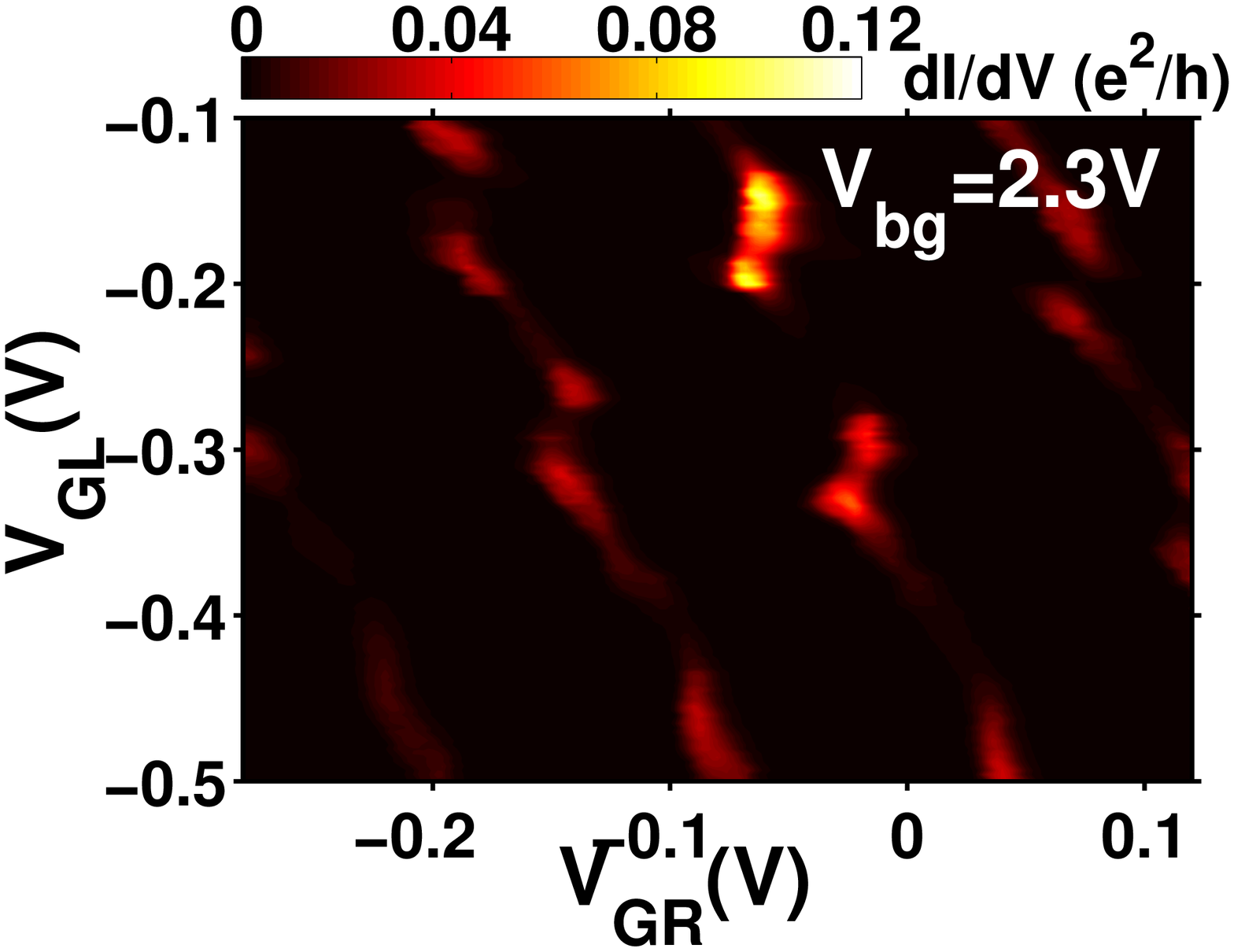}} %
\subfigure[] {\includegraphics[width=0.40\columnwidth]{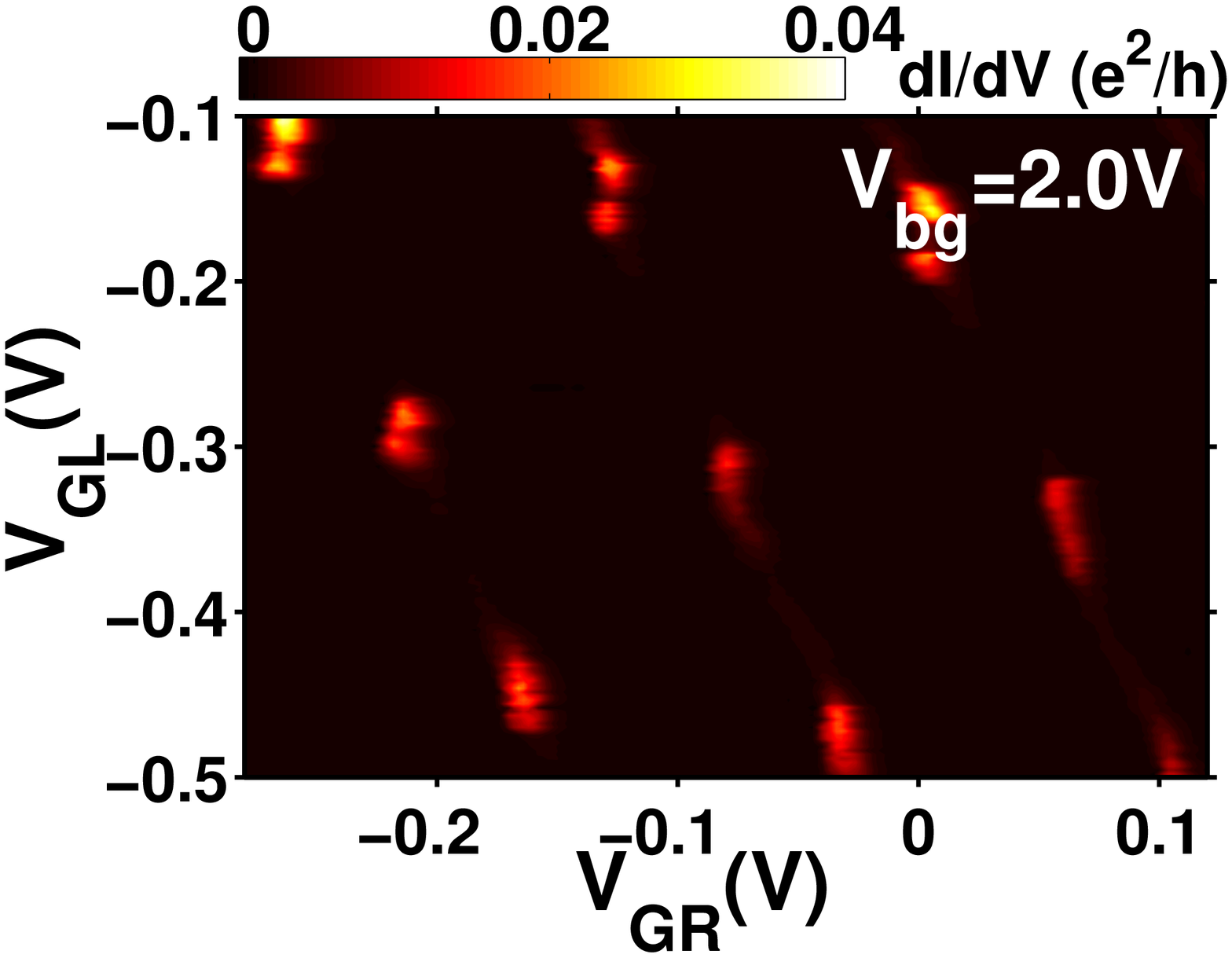}}
\caption{(a)-(c) Colorscale plot of the differential conductance versus
voltage applied on gate~GL ($V_{GL}$) and gate~GR ($V_{GR}$) at $V_{bg}=2.5$
V for different $V_{GL}$ regimes. (c)-(f) Colorscale plot of the
differential conductance versus voltage applied on gate~L ($V_{GL}$) and
gate~R ($V_{GR}$) for different back gate voltage $V_{bg}$. The trend of
interdot tunnel coupling changing from weak to strong can be seen clearly. }
\end{figure}

More quantitative information such as double dot capacitances can be
extracted using a electrostatic model \cite{VanderWiel2003}. First, the
capacitance of the dot to the side gate can be determined from measuring the
size of the honeycomb in Fig. 2(b) as $C_{GL}=e/\Delta V_{GL}$ $\approx 1.27$%
\thinspace aF and $C_{GR}=e/\Delta V_{GR}$ $\approx 1.49$\thinspace aF.
Next, the capacitance ratios can be determined from measuring the size of
the vertices in Fig. 2(b) at finite bias {$V_{sd}=900$\thinspace $\mu $V }as
$\alpha _{L}=|V_{sd}|/\delta V_{GL}=0.029$ and $\alpha _{R}=|V_{sd}|/\delta
V_{GR}\approx 0.035$. Using the relation $C_{GL}/C_{L}=\alpha _{L}$ and $%
C_{GR}/C_{R}=\alpha _{R}$, we can obtain the typical values of dot
capacitances as $C_{L}\approx 44.8$ aF and $C_{R}\approx 44.1$ aF,
respectively. The amount of interdot coupling can be achieved by measuring
the vertices splitting in Fig. 2(c). Assuming the capacitively coupling is
dominant in the weakly coupled dots regime \cite{VanderWiel2003,Marcus2004},
the mutual capacitance between dots is calculated as $C_{m}=\Delta
V_{GL}^{m}C_{GL}C_{R}/e=\Delta V_{GR}^{m}C_{GR}C_{L}/e$ $\approx 9.2$%
\thinspace aF.

It has been expected that opening the interdot constriction by gate voltage
will cause the tunnel coupling to increase exponentially faster than the
capacitive coupling \cite{Kouwenhoven1997}. Fig. 3(a)-(c) represent a
selection of such measurements by holding the same $V_{GR}$ and $V_{bg}$ and
scanning different ranges of $V_{GL}$ between $-0.5$ V to $0.35$ V. An
evolution of conductance pattern indicates that the stability diagram
changes from weak to strong tunneling regimes \cite%
{VanderWiel2003,Marcus2004}. The conductance near the vertices depends on
the relative contributions of the capacitive coupling and tunnel coupling.
For the former, the vertices become a sharpened point, while for the latter,
the vertices become blurred along the edges of the honeycomb cell \cite%
{Graber2006}. In Fig. 3(b), the vertices is not obvious as those in Fig.
3(a), which indicates a stronger tunnel coupling. The results suggest that
two graphene dots are interacting with each other through the large quantum
mechanical tunnel coupling, which is analogous to covalent bonding. We will
analyze it in details below. An increase in interdot coupling also leads to
much larger separation of vertices in Fig. 3(b) \cite{Marcus2004}, and
finally, to a smearing of honeycomb features in Fig. 3(c). In this case, the
double dots behave like a single dot. We note that a similar evolution is
observed for four different values of $V_{bg}$ from $2.5$ V to $2.0$ V at
the same $V_{GL}$ and $V_{GR}$ regimes. Thus the interdot tunnel coupling
could also be changed by $V_{GL}$ or $V_{bg}$. This can be explained by the
fact that the side gates and back gate may influence the central barrier
through the existing capacitances between the gates and the central barrier.

\begin{figure}[tbp]
\subfigure[] {\includegraphics[width=0.4\columnwidth]{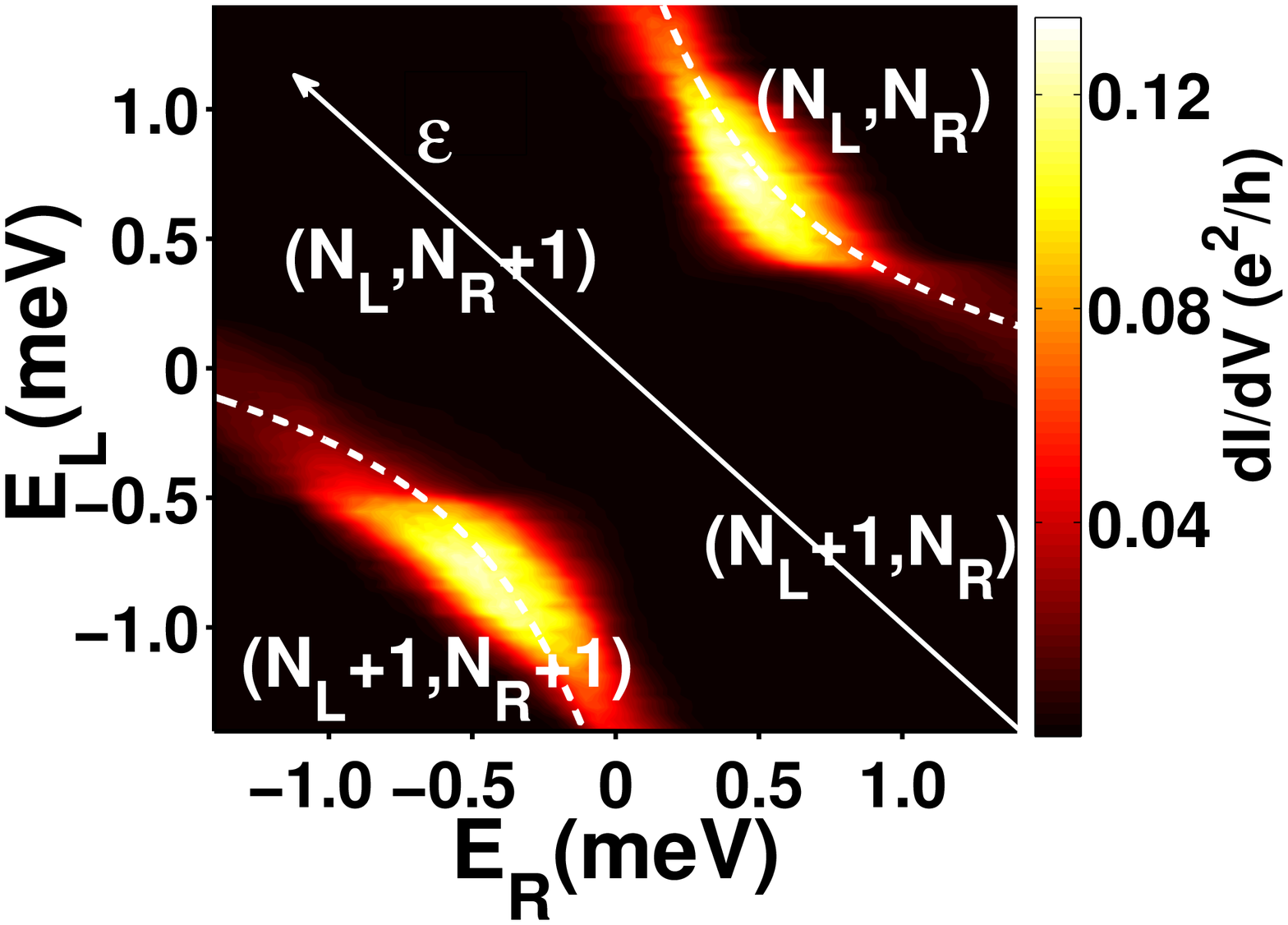}} %
\subfigure[] {\includegraphics[width=0.3\columnwidth]{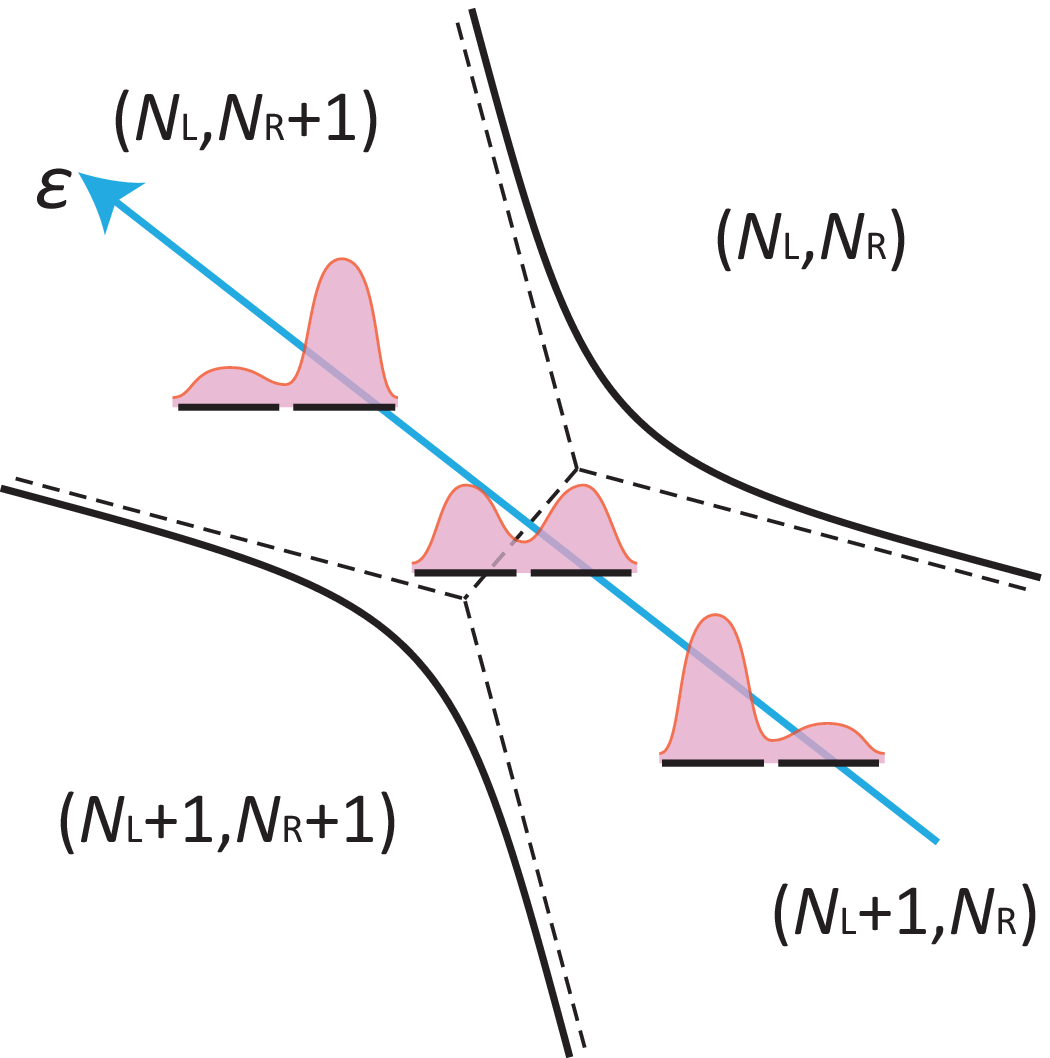}} %
\subfigure[] {\includegraphics[width=0.4\columnwidth]{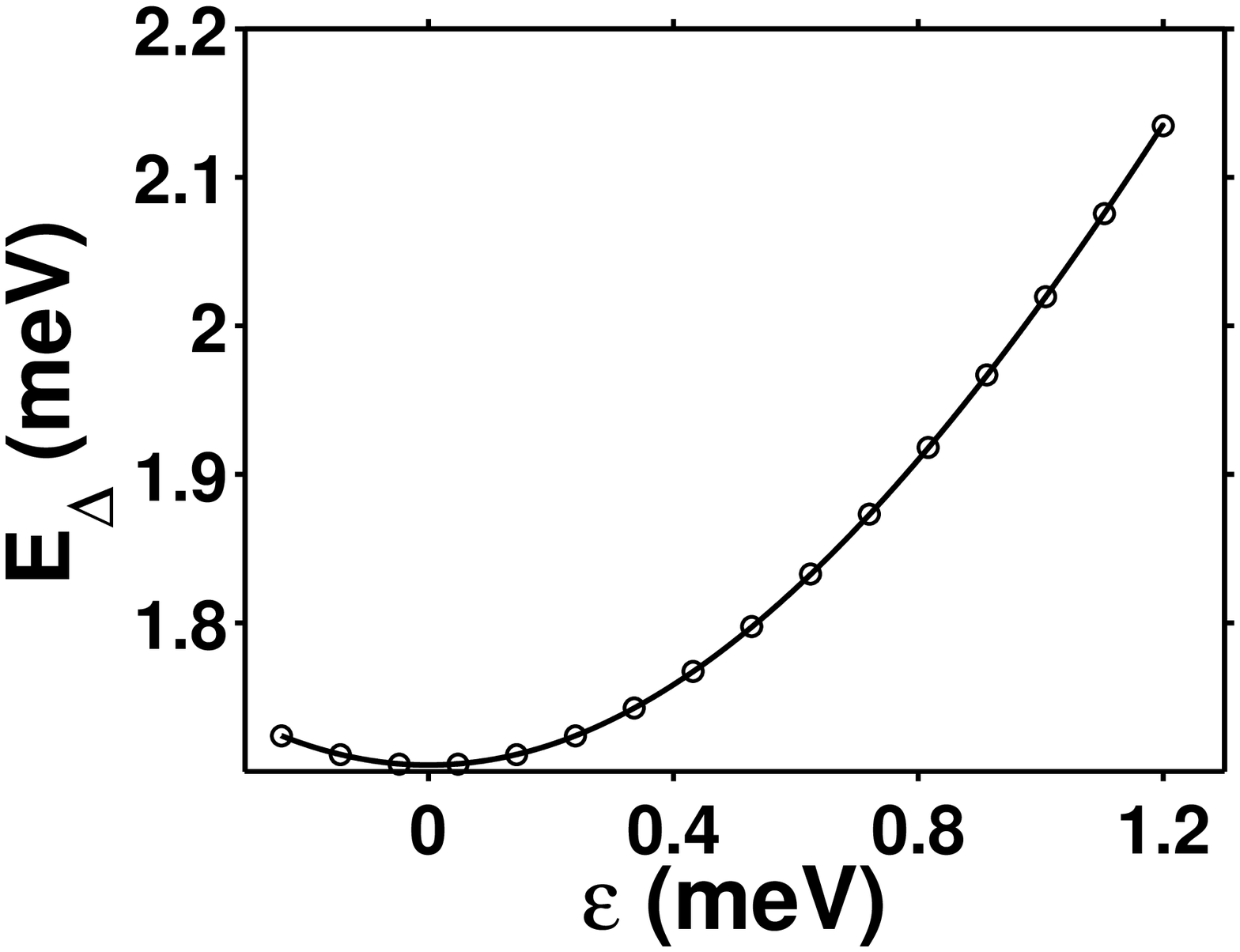}}
\caption{(a) Colorscale plot of the differential conductance versus the
energies of each dot $E_{L}$ and $E_{R}$ at $V_{sd}=20$\thinspace $\protect%
\mu $V near the selected two vertices with dashed lines as guides to the
eye. (b) Schematic of a single anticrossing and the evolution from the state
localized in each dot to a molecule state extending across both dots
\protect\cite{Hatano2005}. (c) $E_{\Delta }$ dependence of the detuning $%
\protect\epsilon =E_{L}-E_{R}$. $E_{\Delta }$ (circles) is measured from the
separation of the two high conductance wings in Fig. 4(a). The line
illustrates a fit of the data to Eq. (1). }
\end{figure}

Having understood the qualitative behavior of the graphene device in the
strong coupling regime, we extract the quantitative properties based on a
quantum model of graphene artificial molecule states \cite%
{Graber2006,Hatano2005}. Here we only take into account the topmost occupied
state in each dot and treat the other electrons as an inert core \cite%
{VanderWiel2003,Golovach2004}. In the case of neglected tunnel coupling, the
nonzero conductance can only occur right at the vertices which are energy
degenerate points as $E(N_{L}+1,N_{R})=E(N_{L},N_{R}+1)$. When an electron
can tunnel coherently between the two dots, the eigenstates of the double
dot system become the superposed states of two well-separated dot states
with the form%
\begin{eqnarray*}
|\Psi _{B}\rangle &=&-\sin \frac{\theta }{2}e^{-\frac{i\varphi }{2}%
}|N_{L}+1,N_{R}\rangle +\cos \frac{\theta }{2}e^{\frac{i\varphi }{2}%
}|N_{L},N_{R}+1\rangle , \\
|\Psi _{A}\rangle &=&\cos \frac{\theta }{2}e^{-\frac{i\varphi }{2}%
}|N_{L}+1,N_{R}\rangle +\sin \frac{\theta }{2}e^{\frac{i\varphi }{2}%
}|N_{L},N_{R}+1\rangle ,
\end{eqnarray*}%
where $\theta =\arctan (\frac{2t}{\epsilon })$, $\epsilon =E_{L}-E_{R}$, $%
E_{L}$ and $E_{R}$ are the energies of state $|N_{L}+1,N_{R}\rangle $ and $%
|N_{L},N_{R}+1\rangle $, respectively. Thus $|\Psi _{B}\rangle $ and $|\Psi
_{A}\rangle $ are the bonding and anti-bonding state in terms of the
uncoupled dot, and the energy difference between these two states can be
expressed by \cite{Graber2006}
\begin{equation}
E_{\Delta }=U^{\prime }+\sqrt{\epsilon ^{2}+(2t)^{2}}.
\end{equation}%
Here $U^{\prime }$ $=\frac{2e^{2}C_{m}}{C_{L}C_{R}-C_{m}^{2}}$ is the
contribution from electrostatic coupling between dots \cite{Bruder2000}.

Provided that the graphene double-dot molecule eigenstate $|\Psi \rangle $
participates in the transport process, sequential tunneling is also possible
along the honeycomb edges \cite{Graber2006}. In Fig. 4(a), a colorscale plot
of the differential conductance is shown at $V_{sd}=20\,\mu $V in the
vicinity of a vertex. As expected the visible conductance is observed at
both the position of the vertex and the honeycomb edges extending from the
vertex. Fig. 4(b) shows a fit of the energy difference $E_{\Delta }$ from
the measured mount of splitting of the positions of the differential
conductance resonance peak in the $\epsilon $-direction. Here we use $%
\epsilon =E_{L}-E_{R}=e\alpha _{L}V_{GL}-e\alpha _{R}V_{GR}$ to translate
the gate voltage detuning $V_{GL}-V_{GR}$ with the conversion factors $%
\alpha _{L}$ and $\alpha _{R}$ determined above. The fitting with Eq. (1)
yields the values of tunnel coupling strength $t$ $\approx 727$\thinspace $%
\mu $eV and $U^{\prime }\approx 209\,\mu $eV. Similar measurements have been
performed in a carbon nanotube double dots with $t\approx 358$ $\mu $eV and $%
U^{\prime }\approx 16$ $\mu $eV~\cite{Graber2006} and semiconductor double
dots with $t\approx 80$ $\mu $eV and $U^{\prime }\approx 175$ $\mu $eV~\cite%
{Hatano2005}. The fact that the tunnel coupling $t\ $is dominant than
capacitive coupling $U^{\prime }$ implies the interdot tunnel barrier in this
etched graphene double dot is much more transparent than those gated carbon
nanotube or semiconductor double dot.

Finally, we discuss the relevance of graphene double dot device for
implementing a quantum gate and quantum entanglement of coupled electron
spins. A $\sqrt{\text{SWAP}}$ operation has already been demonstrated in a
semiconductor double dot system using the fast control of exchange coupling $%
J$ \cite{Petta2005}. The operation time $\tau $ is about $180$ ps for $%
J\approx 0.04$ meV corresponding to $t$ $\approx 0.16$ meV. In the present
graphene device, we have obtained much larger $t\approx 0.72$ meV and the
estimated $\tau \approx 50$ ps is much shorter than the predicted
decoherence time ($\mu $S) \cite{Loss2009-2}. The results indicate the
ability to carry out two-electron spin operations in nanosecond timescales
on a graphene device, four times faster than perviously shown for
semiconductor double dot.

In conclusion, we have measured a graphene double quantum dot with multiple
electrostatic gates and observed the transport pattern evolution in
different gate configurations. This way offers us a method to identify the
molecular states as a quantum-mechanical superposition of double dot and
measure the contribution of the interdot tunneling to the splitting of the
differential conductance vertex. The precisely extracted values of interdot
tunnel coupling for this system is much larger than those in previously
reported semiconductor device. These short operation times due to large
tunneling strength together with the predicted very long coherence times
suggest that the requirements for implementing quantum information
processing in graphene nanodevice are within reach.

This work was supported by the National Basic Research Program of China
(Grants No. 2009CB929600), the National Natural Science Foundation of China
(Grants No. 10804104, No. 10874163, No. 10934006, No. 11074243).

%%%%%%%%%%%%%%%%%%%%%%%%%%%%%%%%%%%%%%%%%%%%%%%%%%%%%%%%%%%%%%%%%%%%%%%%%%%%%%%%%%

\end{document}